# Was There A Big Bang?

Robert K. Soberman,[1] Maurice Dubin[2]

Abstract: The big bang hypothesis is widely accepted despite numerous physics conflicts.  It rests upon two experimental supports, galactic red shift and the cosmic microwave background.  Both are produced by dark matter, shown here to be hydrogen dominated aggregates with a few percent of helium nodules.  Scattering from these non-radiating intergalactic masses produce a red shift that normally correlates with distance.  Warmed by our galaxy to an Eigenvalue of 2.735 K, drawn near the Earth, these bodies, kept cold by ablation, resonance radiate the Planckian microwave signal.  Several tests are proposed that will distinguish between this model and the big bang.

**Introduction**

Although predicted from general relativity, the concept of the universe emerging from a singularity was sufficiently bizarre that the term "big bang" was applied in derision.   Although Hubble did not believe the galactic red shifts were due to Doppler recession, he estimated the recession velocities if that were the cause (Hubble & Humason, 1931).  Most remained skeptical of an expanding universe until the discovery of a "cosmic microwave background" by Penzias and Wilson (1965) assumed to be the cooled radiation from the big bang predicted by Alpher, Gamow and Herman (1967).  There were two serious problems with that assumption; i.e., the spatial uniformity (to about one part in $10^5$) and the Planckian curve at 2.735 K (Mather et al., 1990).  The uniformity, in contradiction to the observed universe required an assumed period of inflation for which no physical explanation exists.  In quantum mechanics a Planckian blackbody radiation distribution results only from assemblages of baryonic matter at the appropriate temperature.  As hypothesized big bang remnants are not baryonic ensembles, a Planck distribution is a significant anomaly and no explanation exists how it results from cooling of exploded inflated remnants.  Cosmologists simply assume that the cooled big bang remnant radiates a Planckian (Smoot & Scott, 1996).

 To conform to the observed universe, data from the Differential Microwave Radiometer (DMR) carried by the Cosmic Background Explorer (COBE) satellite was subjected to detailed analysis from which micro-Kelvin variations were found (Smoot, 1992).  Despite bearing no relation to the visible universe these micro-variations were assumed the forerunners of galaxies, stars, etc. (Smoot & Scott, 1996).  Finer scale measuring devices such as the Wilkenson Microwave Anisotropy Probe (Bennett et al., 2003) have been deployed.  Temperature (Hinshaw et al., 2007) and polarization analyses (Page et al., 2007) from WMAP found foreground (Milky Way) influences. However, WMAP produced no better association to the universe.

Despite numerous theoretical (cosmological) interpretations, the big bang is supported by only two experimental observations, the galactic red shift and the cosmic microwave background (CMB).  In what follows, we offer alternative explanations for both, supported by experimental evidence.  Bereft of these two supporting pieces of evidence, the big bang hypothesis should collapse.  Any hypothesis worthy of consideration should offer predictions that allow choice between it and competitor(s). This model concludes with analytical and experimental predictions, the results of which should contradict the big bang hypothesis.

**Dark Matter**

The two experimental supports for the big bang, the CMB and the galactic red shift are products of the dark matter pervading the universe. To explain, it is first necessary to describe the baryonic nature of dark matter, its discovery and what is known of its characteristics. As the baryonic nature of the dark matter is key to understanding its effects, multiple experiments supporting statements herein are cited.

Although interstellar meteors were predicted from telescopic observations of the local spiral arm in the middle of the twentieth century (Öpik, 1950), discovery was delayed by their near invisibility. A reexamination of data from the Asteroid/Meteoroid Experiment carried by the Pioneer 10 and 11 interplanetary spacecraft revealed this population (Dubin & Soberman, 1991). Dubbed cosmoids (a contraction of cosmic meteoroid), these baryonic masses, composed predominantly of frozen hydrogen with a few percent helium form the dark matter of the universe. Below we describe the aggregates and the characteristics that allowed earlier detection escape. This extremely dark fragile population agglomerates in the near absolute zero cold and almost forceless space between galaxies from material expelled in stellar winds. Little, if any radiation is emitted at that temperature, hence its invisibility. Energy loss of photons Mie (1908) scattered predominantly forward from this population results in the intergalactic red shifts that normally correlate with distance. Drawn gravitationally into our galaxy, the solar system and the immediate Earth vicinity, heated by the Sun, these ensembles, remaining cold by hydrogen ablation, resonance radiate the 2.735 K Planckian that is erroneously interpreted (Alpha et al. 1967) as the "big bang cosmic microwave background."

From the reexamination of the 283 events recorded by the Pioneer 10/11 Asteroid /Meteoroid Experiment (Dubin & Soberman, 1991), supported by other independent experiments, characteristics of this population emerged. These allowed earlier detection escape.
1. inability to produce photographic (visual) meteors
2. the ionization trail is unable to return measurable backscatter echoes at normal radar frequencies
3. highly unlikely to damage spacecraft surfaces
4. atmospheric interaction is undetectable by traditional instruments or if noticed, attributable to established environmental phenomena.

Characteristics 1, 2, and 3 are the consequence of the extreme fragility of this population.
As stated above, independent experiments that support claims are cited. The results from the dust experiments aboard the two Helios spacecraft that explored the inner solar system lend such support (Grün et al. 1980). Each of the two rotating Helios spacecraft carried two dust detectors, one aimed into the ecliptic and the other at higher solar latitudes. To protect the ecliptic detector from direct sunlight, the opening was covered with a thin (3.75 micrometer) coated plastic shield. An unpredicted "eccentric" population, so called for the high eccentricities of their orbits ("even hyperbolic could not be excluded") was measured only by the out of the ecliptic detectors. The investigators concluded that this population could not penetrate the thin protective shields (Grün et al. 1980).

Addressing both the fragile nature of cosmoids and the inability to yield reflections at normal radar frequencies, it was not until high power 2 MHz (very low frequency VLF) radar built for military long range communication was directed at meteors that Olsson-Steel and Elford (1987) measured echoes from a population that produced coherent electron trails at altitudes above 120 kilometers. This population has an altitude distribution completely different from traditional radio meteors measured at much higher frequencies. The peak of the altitude distribution is near 110 km, with meteors being detected to 140 km. The high altitude data implies that the number of meteors ablating continues to rise with increasing height indicating, as the experimenters contended, an extremely fragile structure (Olsson-Steel & Elford 1987). The height distribution for classical radio meteors measured at very high frequency (VHF) peaks at about 90 Km where the air is about 100 times denser. Very few classical VHF radio meteors are detected above 105 kilometers. When calculated by classical solid meteoroid analysis, these high altitude 2 MHz echo returns yield impossibly low densities ($< 10^{-2} g/cm^3$). The computed low density can only be understood as the result of the atmospheric interaction of a freshly dispersed cloud of particles. Further, as reported by the investigators (Olsson-Steel & Elford 1987), this population dominated the meteoroid flux by at least two orders of magnitude.

The interstellar origin of these meteors was clearly identified by Taylor, Baggaley and Steel (1996) using time lags between spaced VLF radar receivers. They reported approximately one percent of over 350,000 echo returns with velocities exceeding 100 km/s. The vector addition of the solar directed escape at 1 AU [42 km/s], the Earth's (near perpendicular) orbital velocity [30 km/s] and the Earth's escape [11 km/s] gives the interstellar cosmoids a terrestrial accretion velocity somewhat greater than 62 Km/s. Taylor, Baggaley and Steel (1996) used the higher limit of 100 km/s to establish interstellar hyperbolic orbit beyond any error related ambiguity. Their 100 km/s cutoff (> 11 times the interstellar cosmoid vector velocity) means that a much larger fraction (i.e., most) measured at this very low frequency originate outside the solar system.

Characteristic 1, inability to produce visual or photographic meteors is a consequence of dispersion of the fragile cosmoids high in the atmosphere where the low light levels resulting are difficult to measure. We note just one example where a faint flash was recorded at 95 Km altitude by a Space Shuttle camera (Boeck, et al. 1992).

Characteristic 4, relating to atmospheric phenomena is peripheral to this paper. It is treated extensively elsewhere (Soberman & Dubin 2001).

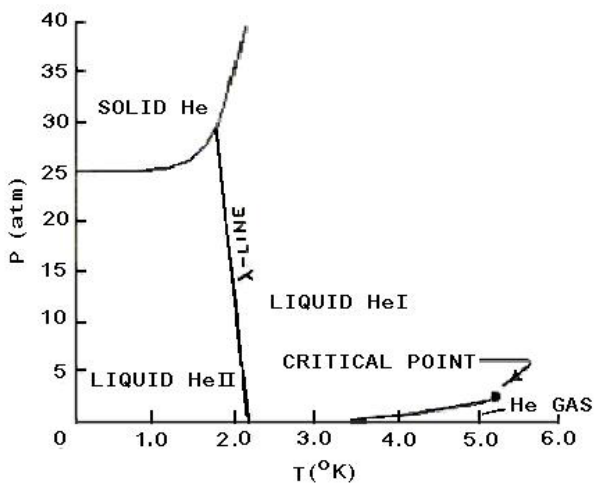

**Figure 1** - Pressure temperature phase diagram for $^4$He [Source: Keller 1969].

Dark matter (cosmoids) are normal baryonic matter composed predominantly of fragile aggregates of frozen hydrogen; the most ubiquitous element in the universe. They are reported to dominate the meteoroid complex by more than two orders of magnitude (Olsson-Steel & Elford 1987). Comet-like, but orders of magnitude smaller, they have an extremely low albedo (2 to 4% estimated) that makes them near invisible until they explosively disrupt (see below). Cosmoids are similar to first apparition comets, unmodified by solar heating. Such a near completely absorbing surface was observed on the nucleus of Comet Halley (Keller, et al. 1986; Sagdeev, et al. 1986) where earlier solar encounters likely increased its albedo. It took the near sacrificial space missions of Giotto, Vega 1 and 2 to get close enough to observe the comet's nucleus.

Predominantly hydrogen with only a few percent of helium, the cold heavier super-conductive liquid helium settles to form nodules in the aggregates with the poorly conducting hydrogen forming a frozen mantle. Warming produces a temperature near 3 K. In warming the helium must transition through state and phase changes as noted in a text devoted entirely to the properties of helium (Keller 1969). The liquid helium must first transition from the super-conducting super-fluid He II to He I. In pure helium this transition occurs at 2.2 K (see Fig. 1). Phase transition temperatures normally increase in interacting materials and the H-He agglomerate undoubtedly creates its own transition point. As can be seen from figure 1, at this temperature helium evaporation occurs. It should be noted that the phase diagram shown is for pure $^4$He. The interacting hydrogen modifies that temperature. During this process the super-fluid, super-conducting helium liquid conducts the heat quickly if near the surface and turns the entire body into a uniform temperature heat sink. This sets an isothermal condition at which cosmoids must remain during galactic approach and entry. That temperature, fixed between the phase transition for helium and its evaporation from hydrogen dominated agglomerates in near perfect vacuum, is the measured 2.735 K Planckian (Mather, et al. 1990).

Agglomerated in intergalactic space, from stellar wind ejecta, at a radiative equilibrium temperature near zero Kelvin, the hydrogen and helium are solid and liquid respectively. The fluffy porous frozen hydrogen structure provides ample space for super fluid super conducting liquid helium (see Figure 1) nodules to accumulate. At that temperature (~ 0 K) cosmoids act as "cold traps" capturing all atoms, molecules, grains and flakes that come in relative low velocity contact. While the gravitational force of the low mass bodies is small, van der Waal, vacuum cold weld adhesion, surface tension and wetting forces likely provide sufficient bonding strength in the absence of disruptive forces until the mass becomes appreciable and they begin to act as gravitational potential wells. Inherent is the absence of any (big bang) time constraint on the low relative velocity formation of these aggregates. The fluffy structure that aids agglomeration also provides thermal insulation to keep the interior cold; maintaining liquid helium when exposed for short times to stellar radiation. Ablation of the surrounding hydrogen (as with comets) keeps the helium from evaporating during prolonged close stellar encounters.

**Transit to the Earth's vicinity**

Like the nuclei of comets, cosmoids are extremely dark (i.e. have extremely low albedo) hence very high emissivity and absorbtivity. The experimental evidence for this is their pronged hidden nature. As to their high emissivity, experimental evidence will be cited later. At great distance from galaxies, the dark matter (cosmoid) population remains stable at near absolute zero radiative equilibrium temperature. Hence there is virtually no thermal radiation. Near a galaxy such as the Milky Way cosmoids absorb ambient starlight and cosmic rays and begin to warm. Eddington (1926) calculated with only the energy of relatively nearby stars to warm them, the radiative equilibrium temperature of diffuse matter in interstellar space should be about 3.18 K.

During warming the super-fluid, super-conducting helium liquid conducts the heat and turns the entire body into a uniform temperature heat sink. The helium nodules set the isothermal condition at which cosmoids must remain during galactic approach and entry. Fixed between the phase transition for helium and its evaporation from hydrogen dominated agglomerates in near perfect vacuum, that temperature is 2.735 K. At that temperature, with high emissivity, cosmoids are in thermal resonance, absorbing light and cosmic rays from the galaxy and emitting a near ideal blackbody Planck continuum. Radiative equilibrium cannot be achieved until cosmoids are close to a galaxy. As an approximation, consider that cosmoids are sufficiently close to the Milky Way Galaxy for it to be considered a two-dimensional infinite radiation source. We choose this as measurement is more readily made out of the galactic plane. Using the Stefan-Boltzmann relation, radiative equilibrium can be written:

$$\frac{L}{\pi R^2} \pi a^2 [1 - A] = 4 \pi a^2 \sigma T^4 \qquad (1)$$

where L is the galactic luminosity; a the radius of the cosmoid; A the albedo and T the temperature. Radiative equilibrium is thus seen as independent of body size. Inserting values for the Milky Way, $L = 10^{10} L_s$ (solar luminosities); R = 15 kiloparsecs and [1 - A] near unity; the radiative equilibrium temperature that cosmoids of all sizes achieve close to our galaxy is approximately 2.3 K. With a factor of 2 increase for the energy deposition of charged cosmic ray particles, a radiation equilibrium temperature of approximately 2.7 K results. This should also be about the same equilibrium temperature inside a typical portion of the galactic disk far from the nucleus.

On close galactic approach, having warmed to about 3 K they reach the Eigenvalue thermal plateau that can be sustained by hydrogen ablation even close to a star. That temperature cannot be exceeded until all of the $^4$He evaporates as seen from the phase diagram (Fig. 1). A population of near invisible objects, in and near the galaxy, absorbing galactic radiation and radiating a thermal continuum surrounds any disk star that is far from the galactic nucleus. This Planckian blackbody distribution is that of an Eigenvalue plateau temperature of 2.735 K set by the physical state of the significant constituents.

The agreement of the equilibrium environmental radiation temperature with the measured value (Mather et al. 1990) is not fortuitous in spite of the approximations used. That equilibrium temperature should not be significantly different from the transition temperature. If it were lower, then the cosmoids would not warm to the isotherm plateau, whereas, if it were significantly higher, then in time absorbed energy would cause the helium to evaporate destructively. It is noteworthy, however, that even a large increase in the radiation environment on approach to a star should not cause the emission temperature to change. Both time and domination of the larger bodies (i.e., the luminance is proportional to the surface area) favor maintaining the isotherm until arriving in the immediate vicinity of a star. The loose agglomeration of ice flakes provides good thermal insulation to the interior of the larger bodies. Surface ablation, as with comets, allows the interior to remain cold even in close proximity to a star.

Exposure to the prolonged high temperature stellar radiation environment at the surface of larger cosmoids and throughout the smaller bodies overcomes the thermal energy plateau with consequent helium evaporation. Within a few AU this radiative warming causes surface ablation and comet like jetting leading to the complete disruption of the smallest bodies as recorded in the bright flashes observed from Pioneer 10/11 (Dubin & Soberman 1991). The super conductivity of the helium requires that the phase change occur throughout the nodules. Hence destruction is explosive as measured in the microsecond rise times of the 283 events recorded by the Pioneer 10/11 Asteroid/Meteoroid Experiment (Dubin & Soberman 1991). The photometric data from the Pioneer 10/11 Asteroid/Meteoroid Experiment telescopes showed that sunlight scattered from the explosive dispersion of near wavelength sized micro particles forms the Zodiacal Cloud that surrounds our Sun (Dubin & Soberman 1991).

Local dark matter (cosmoids) in radiation equilibrium dominate measurement of the microwave background from Earth or orbit. The aggregates absorb sunlight and cosmic rays and reemit the Planck thermal Eigenvalue. To remain in equilibrium they must reemit substantial amounts of microwaves as measured from the COBE satellite (Mather et al. 1990). The Earth's extended shadow (umbra and penumbra) shields approaching cosmoids from solar radiation. Like comets they stay near 3 K even at the near 300 K local solar radiation temperature by shedding outer layers, with the consequent loss of the smaller bodies.

The size of cosmoids in the solar system was estimated from the measured Pioneer 10/11 Asteroid/Meteoroid Experiment brightness of the disrupted population (Dubin & Soberman 1991). It extends from nanometers to meter size with a distribution function that decreases two orders of magnitude for each decade of size increase. It was pointed out earlier that the results of the Helios borne dust detectors showed that this population is unlikely to damage spacecraft, as it could not penetrate a very thin plastic shield (Grün et al. 1980).

**Micro-Kelvin variants**

If the 2.735 K microwave radiation is, as we contend, resonance emission from solar heated cosmoids in the immediate Earth vicinity, what then produces the observed micro-Kelvin variations found after meticulous search in the COBE results and interpreted (Smoot 1992) as the forerunners of the observed universe; despite bearing no such semblance? Evidence for interaction between dust clouds and galactic spiral arms was found in the Spitzer Wide-Area Infrared Extragalactic Legacy Survey (SWIRE) that used the cryogenically cooled Spitzer Infrared Telescope Facility (SIRTF). Observations showed starlight in galactic arms masked by dust clouds (Rowan-Robinson, et al. 2005). Meteoric matter contains traces of elements heavier than helium. While the bulk of the approaching cosmoids would be at the equilibrium temperature, these could be warmed slightly by above average radiation. The spiral-armed structure of the Milky Way presents such hotter portions. Compton scatter of energetic electrons and its inverse; the Sunyaev-Zel'dovich effect (Sunyaev & Zel'dovich 1970) in the spiral arms may also modify the Eigenvalue temperature of the dark matter entering the galaxy. Figure 2 is a plane view of portions of the galactic spiral arms (black) as measured with 21-centimeter (neutral hydrogen line) radio-wavelengths (Seeds 1993). Added are connections (green) obtained from density wave studies (Simonson 1976). To show the micro-Kelvin variants are relatively locally (galaxy) produced, we converted the upper portions of the warmer micro-Kelvin variations (Smoot et al. 1992) attributed to the big bang (Smoot & Scott 1996) to plane polar coordinates (Fig. 3). On this is superposed the galactic configuration (Fig. 2). Allowing for gravitational distortions, blockages, multiple reemissions, plotting and transcription errors, the association (unlike the attribute to the visible universe) is obvious in figure 4.

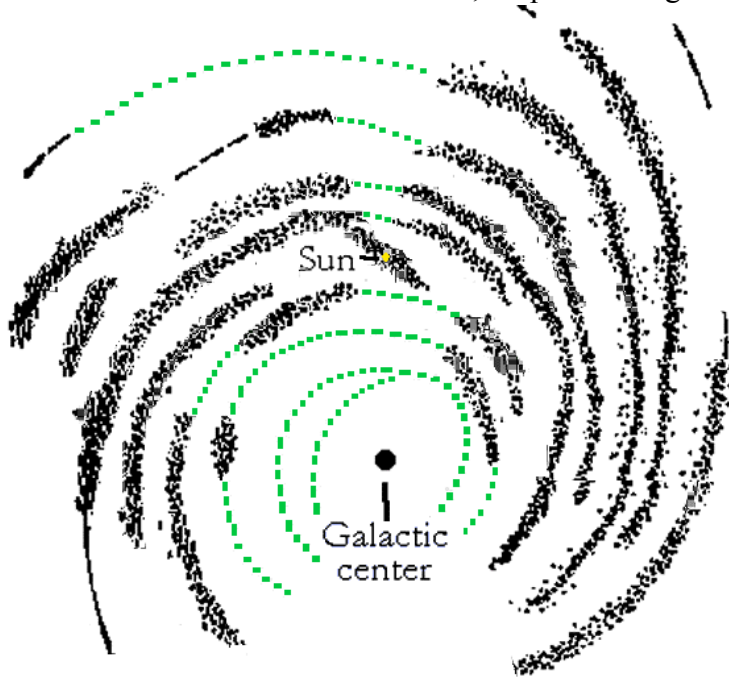

Fig. 2. Spiral arm structure of the Milky Way measured with 21 cm radio-wavelength (black) and density waves (green).

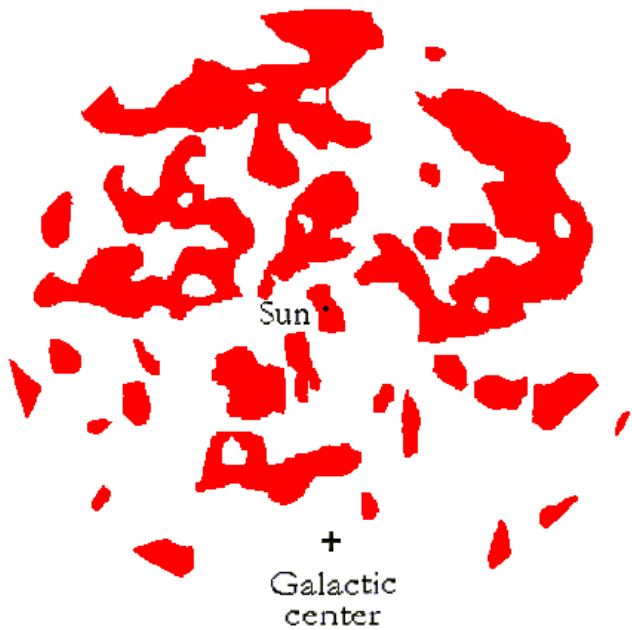

**Fig. 3.** Warm micro-kelvin variations in the microwave background radiation plotted in plane polar coordinates.

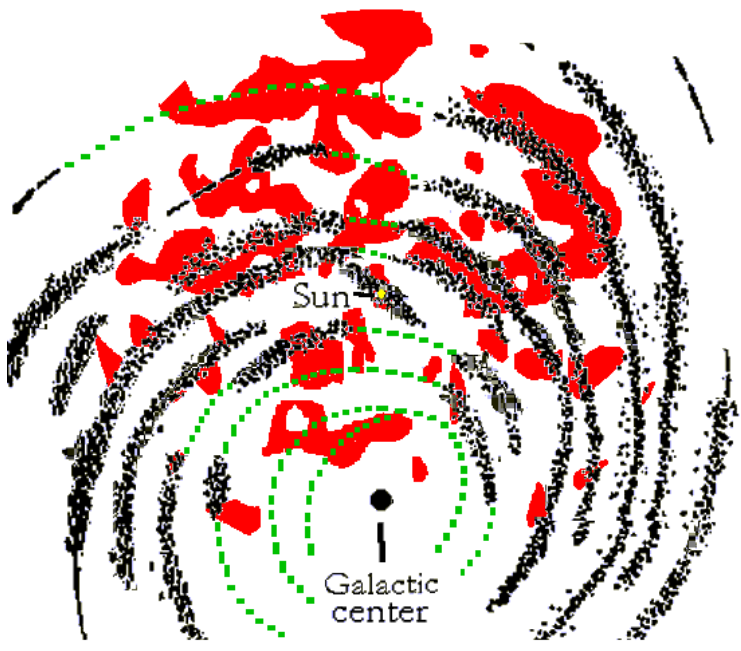

**Fig. 4.** Overlay of figure 2 on figure 3.

**Big Bang Summary**

Having questioned the two supporting experimental pillars (i.e. galactic red shift and microwave background) of the "big bang," we summarize the history of that hypothesis, replete with assumptions and inferences. Sir Fred Hoyle coined the name to deride the idea of the universe emerging from a singularity. The big bang hypothesis evolved from general relativity. Although he did not believe the galactic red shifts were due to Doppler recession, Hubble estimated the recession velocities if they were due to Doppler red shifts (Hubble & Humason 1931). Many remained skeptical of an expanding universe until the discovery of a "cosmic microwave background" by Penzias and Wilson (Penzias & Wilson 1965) that was assumed to be the cooled radiation from the big bang predicted by Alpher, Gamow and Herman (1965). There are two serious problems with that assumption; i.e., the uniformity (to about one part in $10^5$) and the Planckian curve at 2.735 K (Mather et al. 1990). The uniformity, in contradiction to the observed universe required an assumed period of inflation for which there is no physical explanation. Further, in quantum mechanics a Planckian blackbody radiation distribution results only from assemblages of baryonic matter at the appropriate temperature. As hypothesized big bang remnants are not baryonic ensembles, a Planck distribution is a significant anomaly and no explanation exists how it results from cooling of exploded inflated remnants. Cosmologists simply assume that the cooled big bang remnant radiates a Planckian (Smoot & Scott 1996).

To conform to the observed universe, data from the Differential Microwave Radiometer (DMR) carried by the Cosmic Background Explorer (COBE) satellite was subjected to detailed analysis from which micro-Kelvin variations were found (Smoot et al. 1992). Despite bearing no relation to the visible universe these micro-variations were assumed the forerunners of galaxies, stars, etc. (Smoot & Scott 1996). Finer scale measuring devices such as the Wilkenson Microwave Anisotropy Probe (Bennett, et al. 2003) were deployed in the futile search for a better relation. Additional "big bang" data interpretations also persist (Wright et al. 1992; Spergel et al. 2007).

Earlier we pointed out significant discrepancies in the assumption of the galactic red shift resulting from Doppler recession. Arp (1966) has documented hundreds of examples where two or more galaxies apparently interact yet show very different red shifts. More recently, analysis of the "standard candle" brightness of type 1A supernovae gave rise to the assumption of "dark energy," an anti-gravity force acting only on the cosmic level, thus impossible to replicate (Pasqual-Sánchez 1999).

**Conclusions**

It should come as no surprise to the astronomical community that the universe is composed overwhelmingly of hydrogen. The remaining few percent is dominated by helium. For three quarters of a century most of this mass remained cloaked in darkness despite numerous clues to its existence.

The two experimental pillars (red-shift and microwave radiation) on which the big bang rests are shown as consequences of dark matter (cosmoids) in intergalactic, interstellar, solar system and near Earth space. The galactic red-shift results from Mie scatter of transiting intergalactic photons and the microwave foreground radiation from solar heated cosmoids resonance radiating a Planck Eigenvalue set by helium nodules. The micro-Kelvin variants are consequences of interactions with our Milky Way Galaxy.

The import of the "big bang" is apparent in the 2006 Nobel physics prize awards to J. C. Mather and G. F. Smoot. Questioning that hypothesis requires, like any worthwhile physical model, tests that can distinguish between the opposing concepts. Several tests are proposed, the results of which, we predict, will contradict big bang assumptions. Foremost a mixture of hydrogen with a small amount of helium should be cooled to establish that an Eigenvalue plateau exists at 2.735 K. An examination of the intensity of type 1A supernovae radiation versus red shift derived distances will show a small discrepancy from inverse square variance. A microwave receiver aboard an interplanetary spacecraft, beyond 3.5 AU, aimed away from the Sun will hear barely a whisper of the background radiation. Another test involves measuring the red shift produced by the dark matter (cosmoids) lying within one AU of the Sun. If an astronomical source is observed twice, near six months apart, almost 90° from the direction of the Earth's motion, such that the distance difference incorporates a major chord through the Earth's orbital plane, a measurable red shift will be observed.

[1] The Franklin Institute, Philadelphia PA USA (retired). Voorhees, NJ 08043
    Email: rksoberman@yahoo.com
[2] NASA Goddard Space Flight Center, Greenbelt MD USA (retired). Silver Spring, MD 20905.
    Email: mdubin@aol.com.